\documentclass[nofootinbib,11pt,showpacs]{revtex4}

\usepackage{bm,amsmath,amssymb}
\usepackage[dvips]{graphicx}

\begin{document}
\title{\Large Effect of a Weak Electromagnetic Field on Particle Acceleration\\ by a Rotating Black Hole}

\hfill{OCU-PHYS 358, AP-GR 94, RUP-11-7, YITP-11-92}

\pacs{04.70.-s, 04.70.Bw, 97.60.Lf}

\author{$^{1}$Takahisa Igata}
\email{igata@sci.osaka-cu.ac.jp}
\author{$^{2}$Tomohiro Harada}
\email{harada@rikkyo.ac.jp}
\author{$^{3}$Masashi Kimura}
\email{mkimura@yukawa.kyoto-u.ac.jp}

\affiliation{
$^{1}$
Department of Mathematics and Physics,
Graduate School of Science, Osaka City University,
Osaka 558-8585, Japan\\
$^{2}$
Department of Physics, Rikkyo University, Toshima, Tokyo 175-8501, Japan\\
$^{3}$
Yukawa Institute for Theoretical Physics, Kyoto 606-8502, Japan}

\begin{abstract}
We study high energy charged particle collisions near the horizon in an electromagnetic field around a rotating black hole and reveal the condition of the fine-tuning to obtain arbitrarily large center-of-mass (CM) energy. We demonstrate that the CM energy can be arbitrarily large as the uniformly magnetized rotating black hole arbitrarily approaches maximal rotation under the situation that a charged particle plunges from the innermost stable circular orbit (ISCO) and collides with another particle near the horizon. Recently, Frolov [Phys.\ Rev.\ D {\bf 85}, 024020 (2012)] proposed that the CM energy can be arbitrarily high if the magnetic field is arbitrarily strong, when a particle collides with a charged particle orbiting the ISCO with finite energy near the horizon of a uniformly magnetized Schwarzschild black hole. We show that the charged particle orbiting the ISCO around a spinning black hole needs arbitrarily high energy in the strong field limit. This suggests that Frolov's process is unstable against the black hole spin. Nevertheless, we see that magnetic fields may substantially promote the capability of rotating black holes as particle accelerators in astrophysical situations.
\end{abstract}

\maketitle

\section{Introduction}
The possibility that black holes can act as particle accelerators has been intensively studied since Ba\~nados, Silk and West~\cite{Banados:2009pr} discovered that the center-of-mass (CM) energy of two colliding particles can be arbitrarily high if the collision occurs in the vicinity of the horizon of a nearly maximally rotating black hole and if the angular momentum of either of the particles is fine-tuned. Although the effects of self-force might prevent the CM energy from being arbitrarily high~\cite{Berti:2009bk, Jacobson:2009zg, Kimura:2010qy}, it would also be reasonable to assume that the CM energy is still high enough to be of physical interest if the mass-ratio of the particle to the black hole is very small~\cite{Kimura:2010qy, Harada:2011pg}. This process was generalized to most general geodesic particles~\cite{Grib:2010zs, Grib:2010xj, Harada:2010yv, Harada:2011xz}. In particular, it was revealed~\cite{Harada:2010yv} that the fine-tuning of the angular momentum is naturally realized for the particle orbiting an innermost stable circular orbit (ISCO), which is important in astrophysical contexts. In fact, to obtain an arbitrarily high CM energy, the particle motion does not need to be geodesic. It was found~\cite{Zaslavskii:2010aw, Zhu:2011ae} that the CM energy can be arbitrarily high for the collision of charged particles near the horizon of charged black holes and charged rotating black holes, where the fine-tuning of the angular momentum is naturally realized for the particle orbiting the ISCO if it exists.

Magnetic fields prevail in astrophysical compact objects. Although the strength of magnetic fields at the horizon radius of the black hole has never been directly observed, it is estimated to be $\sim 10^{8}$ Gauss for stellar mass black holes and $\sim 10^{4}$ Gauss for supermassive black holes using the relation based on the energy equipartition in accretion flows, the variability plane and the observed characteristic frequency of the X-ray variability~\cite{Piotrovich:2010aq}. These values are fairly strong but still much weaker than the critical one
\begin{equation}
B_\mathrm{max}=c^{4}G^{-3/2}M^{-1}\sim 10^{19}
\left(\frac{M}{M_{\odot}}\right)^{-1}~\mathrm{Gauss},
\end{equation}
above which the self-gravity of magnetic fields directly affects the spacetime geometry around the horizon of the black hole of mass $M$. On the other hand, as we will see later, the effects of the magnetic fields to the particle motion appear with the nondimensional parameter $b$ defined by
\begin{equation}
b=\frac{qBGM}{mc^{4}}\sim8.6\times10^{10}
\left(\frac{q}{e}\right)
\left(\frac{m}{m_{\mathrm e}}\right)^{-1}
\left(\frac{B}{10^{8}\,\mathrm{Gauss}}\right)
\left(\frac{M}{M_{\odot}}\right),
\label{eq:b}
\end{equation}
where $q$ and $m$ are the charge and the rest mass of the particle, respectively. Since $q/m$ is very large for elementary charged particles, the value of $|b|$ is extremely large. In fact, the effects of magnetic fields to the motion of elementary charged particles can never be neglected and the limit $|b|\to \infty$ is expected to give an excellent approximation. The ISCO of charged particles around a weakly magnetized rotating black hole was analyzed in Ref.~\cite{Aliev:2002nw} and more general motion of charged particles around a weakly magnetized Schwarzschild black hole was studied in Ref.~\cite{Frolov:2010mi}.

Recently, Frolov~\cite{Frolov:2011ea} found that when two particles collide near the horizon of a weakly magnetized Schwarzschild black hole, one of which is charged and orbiting the ISCO with a finite Killing energy, the CM energy can be arbitrarily high if the magnetic field is {\it arbitrarily strong}. This phenomenon might be connected to the fact that the radius of the ISCO approaches that of the event horizon in the limit $|b|\to \infty$. This phenomenon should be distinguished from the Ba\~nados-Silk-West (BSW) process in the respect that to obtain an arbitrarily high CM energy Frolov's process needs to presuppose an arbitrarily strong physical field at the beginning. On the other hand, Frolov's process might be relevant to astrophysics because $|b|$ estimated in astrophysical black holes is extremely large.

The aim of this paper is two-fold: to develop the general mathematical framework to obtain the CM energy of non-geodesic particles and, in particular, charged particles in a test electromagnetic field around a Kerr black hole and to identify the effects of test magnetic fields for the charged particles around Schwarzschild and Kerr black holes with a special focus on the CM energy for the collision of a charged particle orbiting the ISCO with a generic neutral counterpart. We then naturally embed Frolov's scenario and Ba\~nados, Silk and West's scenario into the particle acceleration by a weakly magnetized Kerr black hole.

This paper is organized as follows. In the next section, we develop the mathematical framework to obtain the CM energy of non-geodesic particle collisions. In Sec.~\ref{sec:3}, we consider charged particle motion in a uniform magnetic field around a Kerr black hole. In particular, we focus on the ISCO of a charged particle and demonstrate that the radius of the ISCO approaches the horizon radius as the uniform magnetic field is strengthened. In Sec.~\ref{sec:4}, to discuss the effects of a uniform magnetic field on the BSW process and of rotation of the black hole on Frolov's process, we evaluate the CM energy of charged particle collisions around a magnetized Kerr black hole. The final section presents summary and discussion. We use the sign convention $-+++$ for the metric, and units in which $c=G=1$.

\section{General formula for the center-of-mass energy of a charged particle collision}
In this section, we derive the formula for the CM energy of two colliding charged particles in a stationary and asymmetric test electromagnetic field around a Kerr black hole. We discuss the effect of an electromagnetic field on the CM energy and whether the BSW process arises from charged particle collisions.

The metric of the Kerr geometry in the Boyer-Lindquist coordinates is given by
\begin{align}
ds^2
=&-\left(\frac{\Delta-a^2\sin^2\theta}{\Sigma}\right)dt^2
-\frac{2a\sin^2\theta(r^2+a^2-\Delta)}{\Sigma}dtd\phi
\cr
&+\frac{(r^2+a^2)^2-\Delta a^2\sin^2\theta}{\Sigma}\sin^2\theta d\phi^2
+\frac{\Sigma}{\Delta}dr^2+\Sigma d\theta^2, 
\label{eq:metric}
\end{align}
where
\begin{align}
\Sigma&=r^2+a^2\cos^2\theta,\\
\Delta&=r^2+a^2-2Mr,
\end{align}
and the two parameters, $a$ and $M$, denote the spin and the mass of a Kerr black hole, respectively. In the case $a^2 \leq M^2$, the event horizon of the Kerr black hole is located at the $r$-coordinate $r_\mathrm{H} = M +\sqrt{M^2-a^2}$, where $\Delta$ vanishes. As one approaches the event horizon, $r\to r_\mathrm{H}$, her or his coordinate angular velocity $\Omega$, defined by $\Omega=d\phi/dt$, approaches
\begin{align}
\Omega_\mathrm{H}=\frac{a}{r_\mathrm{H}^2+a^2},
\end{align}
which is called the angular velocity of the horizon. Using $\Omega_\mathrm{H}$ we can write the Killing vector that is tangent to the null geodesic generator of the horizon
\begin{align}
\chi^a=\xi^a+\Omega_\mathrm{H}\psi^a,
\end{align}
where $\xi^a=(\partial/\partial t)^a$ is a stationary Killing vector and $\psi^a=(\partial/\partial \phi)^a$ is an axial one.\footnote{The coordinate bases $\xi^a = (\partial /\partial t)^a$ and $\psi^a = (\partial /\partial \psi)^a$ are regular at the horizon $r=r_\mathrm{H}$, while the Boyer-Lindquist coordinates do not cover the horizon. We confirm that by using the Kerr-Schild coordinates as shown in App.~\ref{app:A}.}

We consider the motion of a charged particle in a stationary and axisymmetric Maxwell field around a Kerr black hole. Let $p_a$ be the canonical momentum of a charged particle conjugate to the coordinates. Then the Hamiltonian of charged particle motion is given by
\begin{align}
H=\frac12g^{ab}(p_a-q A_a)(p_b-q A_b),
\end{align}
where $q$ is the charge of a particle and $A^a$ is a stationary and axisymmetric vector potential. Then we have two constants of motion $p_t$ and $p_{\phi}$ because $t$ and $\phi$ are cyclic coordinates due to the assumption of the spacetime symmetries. By the Hamilton equation, the momentum of a charged particle measured by a local observer, $\pi^a=\partial H/\partial p_a$, is of the form
\begin{align}
\pi^a=g^{ab}p_b-qA^a.
\end{align}
Note that we write $\pi^a=mu^a$ for a timelike particle of mass $m$, where $u^a$ is the velocity, and write $\pi^a=\hbar k^a$ for a null particle with the wave vector $k^a$. For simplicity, we restrict our attention to particle motion on the equatorial plane ($\theta=\pi/2$) so that $\pi^{\theta}=0$. By using the normalization condition, $g_{ab}\pi^a\pi^b=-m^2$, where $m=0$ for a null particle, the equation of the radial motion becomes
\begin{align}
\pi^r=\frac{\kappa\sqrt{\mathcal{R}}}{r^2},
\label{eq:pi^r}
\end{align}
where $\kappa=\pm1$ and the function $\mathcal{R}$ is given by
\begin{align}
\mathcal{R}=\mathcal{P}^2-\Delta\mathcal{F},
\end{align}
where
\begin{align}
\mathcal{P}&=(r^2+a^2)(-\pi_t)-a\pi_\phi,\\
\mathcal{F}&=\left(\pi_{\phi}+a\pi_t\right)^2+m^2r^2.
\end{align}

Let us consider the CM energy of two particles $\pi^a_{(s)}$ of mass $m_s$ and charge $q_s$ labeled by $s=1, 2$. Hereafter $p_a^{(s)}$, $\kappa_s$, $\mathcal{R}_s$, $\mathcal{P}_s$, and $\mathcal{F}_s$ represent $p_a$, $\kappa$, $\mathcal{R}$, $\mathcal{P}$, and $\mathcal{F}$ for particle-$s$, respectively. The CM energy $E_\mathrm{cm}$ at a collision point is defined by
\begin{align}
E_\mathrm{cm}^2=-P_aP^a=m_1^2+m_2^2-2g_{ab}\pi_{(1)}^a\pi_{(2)}^b,
\label{eq:definition of Ecm}
\end{align}
where $P^a$ is the total momentum of the two charged particles, which is given by $\pi_{(1)}^a+\pi_{(2)}^a$. Using Eqs.~\eqref{eq:metric} and \eqref{eq:definition of Ecm}, we obtain the CM energy of charged particle collisions in an electromagnetic field around the Kerr black hole
\begin{align}
E_\mathrm{cm}^2
=&m_1^2+m_2^2+\frac{2}{r^2}\left(\frac{\mathcal{P}_1\mathcal{P}_2-\kappa_{1}\sqrt{\mathcal{R}_1}\kappa_{2}\sqrt{\mathcal{R}_2}}{\Delta}
-(\pi^{(1)}_\phi+a\pi_t^{(1)})(\pi^{(2)}_\phi+a\pi_t^{(2)})\right),
\label{eq:Ecm^2}
\end{align}
where we have used the radial equation of motion given in Eq.~\eqref{eq:pi^r}. 

Let us evaluate the CM energy near the horizon in what follows. Equation \eqref{eq:Ecm^2} implies that the necessary condition to make $E_\mathrm{cm}$ infinite is that $\Delta$ is infinitesimal, that is, the collision must occur near the horizon. If $\kappa_1=\kappa_2$ and $\mathcal{R}_s\neq0$ for each $s$, we have
\begin{align}
\lim_{r\to r_\mathrm{H}}\frac{\mathcal{P}_1\mathcal{P}_2-\sqrt{\mathcal{R}_1}\sqrt{\mathcal{R}_2}}{\Delta}
=\lim_{r\to r_\mathrm{H}}\left(\frac{\mathcal{P}_2}{2\mathcal{P}_1}\mathcal{F}_1+\frac{\mathcal{P}_1}{2\mathcal{P}_2}\mathcal{F}_2\right),
\label{eq:l'Hopital}
\end{align}
where we have used l'Hopital's theorem. From Eqs.~\eqref{eq:Ecm^2} and \eqref{eq:l'Hopital}, the CM energy in the horizon limit is given by
\begin{align}
\left(E_\mathrm{cm}^\mathrm{H}\right)^2
=\lim_{r\to r_\mathrm{H}}E_\mathrm{cm}^2
=m_1^2+m_2^2+\frac{1}{r_\mathrm{H}^2}\left(
\left(\mathcal{J}_1^2+m_1^2r_\mathrm{H}^2\right)\frac{\mathcal{I}_2}{\mathcal{I}_1}
+\left(\mathcal{J}_2^2+m_2^2r_\mathrm{H}^2\right)\frac{\mathcal{I}_1}{\mathcal{I}_2}-2\mathcal{J}_1\mathcal{J}_2\right),
\label{eq:preEcm^H}
\end{align}
where we have defined $\mathcal{I}_s$ and $\mathcal{J}_s$ as
\begin{equation}
\mathcal{I}_s=\lim_{r\to r_\mathrm{H}}\left(-\chi^a\pi_a^{(s)}\right)
\end{equation}
and
\begin{equation}
\mathcal{J}_s=\lim_{r\to r_\mathrm{H}}\left(\pi^{(s)}_{\phi}+a\pi_t^{(s)}\right),
\end{equation}
respectively. Rearranging Eq.~\eqref{eq:preEcm^H}, we obtain
\begin{align}
E_\mathrm{cm}^\mathrm{H}
=\sqrt{(m_1+m_2)^2+\frac{(\mathcal{J}_1\mathcal{I}_2-\mathcal{J}_2\mathcal{I}_1)^2
+r_\mathrm{H}^2(m_1\mathcal{I}_2-m_2\mathcal{I}_1)^2}{r_\mathrm{H}^2\mathcal{I}_1\mathcal{I}_2}}.
\label{eq:Ecm^H(gerenal formula)}
\end{align}
This is the general formula for the CM energy of charged particle collisions near the horizon on the equatorial plane in an electromagnetic field. Furthermore, for two timelike charged particles, we can rewrite Eq.~\eqref{eq:Ecm^H(gerenal formula)} in a simpler form as
\begin{align}
\frac{E_\mathrm{cm}^\mathrm{H}}{\sqrt{m_1m_2}}
=\sqrt{\frac{(m_1+m_2)^2}{m_1m_2}
+\frac{(\hat{\mathcal{J}}_1\hat{\mathcal{I}}_2-\hat{\mathcal{J}}_2\hat{\mathcal{I}}_1)^2+r_\mathrm{H}^2(\hat{\mathcal{I}}_2-\hat{\mathcal{I}}_1)^2}{r_\mathrm{H}^2\hat{\mathcal{I}}_1\hat{\mathcal{I}}_2}},
\label{eq:Ecm^H}
\end{align}
where the functions $\hat{\mathcal{I}}_s$ and $\hat{\mathcal{J}}_s$ are defined to be $\mathcal{I}_s/m_s$ and $\mathcal{J}_s/m_s$, respectively. Equation \eqref{eq:Ecm^H(gerenal formula)} or \eqref{eq:Ecm^H} shows that the necessary condition to have arbitrarily large $E_\mathrm{cm}$ at the horizon is that $\mathcal{I}_s$ is arbitrarily close to zero for either $s$. We call the particle satisfying $\mathcal{I}_s=0$ a critical particle, which is defined in Ref.~\cite{Harada:2010yv}.

We assume that the vector potential of a stationary and axisymmetric electromagnetic field is of the form 
\begin{align}
A^a=\alpha\xi^a+\beta\psi^a,
\label{eq:A}
\end{align}
where $\alpha$ and $\beta$ are functions of $r$ and $\theta$ and $A_a$ is a regular vector on and outside the event horizon, i.e., $\alpha$ and $\beta$ take finite values in those regions. In the case of two charged massive particle collisions, the CM energy $E_\mathrm{cm}$ is written in the form of Eq.~\eqref{eq:Ecm^2} with the functions
\begin{equation}
\pi^{(s)}_{\phi}+a\pi_{t}^{(s)}=L_s-aE_s-q_s\left[(r^2+a^2)\beta-a\alpha\right]
\end{equation}
and
\begin{equation}
\mathcal{P}_s=(r^2+a^2)E_s-aL_s+q_s(a\beta-\alpha)\Delta,
\end{equation}
where we have defined $E_s=-p^{(s)}_t$ and $L_s=p_{\phi}^{(s)}$, which are interpreted as conserved energy and angular momentum of particle-$s$, respectively.

The expression of $E_\mathrm{cm}^\mathrm{H}$ in this case includes the functions $\mathcal{I}_s$ and $\mathcal{J}_s$ given by
\begin{equation}
\mathcal{I}_s=E_s-\Omega_\mathrm{H}L_s+q_s\Phi_\mathrm{H},
\label{eq:I_s}
\end{equation}
and
\begin{equation}
\mathcal{J}_s=L_s-aE_s-\frac{q_sa}{\Omega_\mathrm{H}}(\beta-\Omega_\mathrm{H}\alpha),
\end{equation}
respectively, where $\Phi_\mathrm{H}=\lim_{r\to r_\mathrm{H}}\chi^aA_a$ is called the electric potential of the black hole. If we assume that the electromagnetic field vanishes, $A_a=0$, and two colliding particles have the same rest mass $m_0$, Eq.~\eqref{eq:Ecm^H(gerenal formula)} coincides with Eq.~(3.5) in Ref.~\cite{Harada:2010yv},
\begin{align}
\frac{E_\mathrm{cm}^\mathrm{H}}{2m_0}
=\sqrt{1+\frac{4M^2\left[(E_1-\Omega_\mathrm{H}L_1)-(E_2-\Omega_\mathrm{H}L_2)\right]^2+(E_1L_2-E_2L_1)^2}{16M^2(E_1-\Omega_\mathrm{H}L_1)(E_2-\Omega_\mathrm{H}L_2)}}.
\end{align}

\section{Charged particle motion in a uniform magnetic field around a rotating black hole}
\label{sec:3}
In this section, we discuss charged particle motion around a Kerr black hole immersed in a uniform magnetic field. We focus on the ISCO of a charged particle because the motion of a particle plunging from the ISCO to the horizon is one of the fundamental processes.\footnote{In App.~\ref{app:B}, we discuss the condition for a charged particle in a circular orbit to reach the ISCO by emitting its energy and angular momentum.} In particular, we show that the ISCO shifts toward the event horizon due to the effect of the magnetic field, which may cause the fine-tuning of the angular momentum because such fine-tuning occurs when the ISCO radius approaches the horizon radius in the case of a neutral particle collisions.

We specify two functions in Eq.~\eqref{eq:A} as
\begin{align}
\alpha=0,\quad\beta=B/2,
\label{eq:uniform magnetic field}
\end{align}
where $B$ is a constant. The vector potential solves vacuum Maxwell's equation, which was discussed in Ref.~\cite{Wald:1974np}. At large distances from the Kerr black hole, the field strength $F_{ab}=2\nabla_{[a}A_{b]}$ asymptotically approaches the uniform magnetic field of strength $B$, which is parallel to the rotation axis of the black hole. As the first step, we adopt the uniformly magnetized black hole as the background spacetime of charged particle motion\footnote{Note that the present choice of the magnetic field is one of many other possible field configurations, as discussed in Ref.~\cite{Wald:1974np, Petterson:1975sg}.}.

To see the dependence of the ISCO radius on the magnetic field, we discuss the radial motion of a charged particle on the equatorial plane, which is also discussed in Ref.~\cite{Aliev:2002nw, Frolov:2010mi}. For simplicity we omit the index $s$, which will be restored in the next section. From Eq.~\eqref{eq:pi^r}, we have
\begin{align}
\frac12\dot r+V_\mathrm{eff}=0,
\end{align}
where $V_\mathrm{eff}=-\mathcal{R}(r)/(2m^2r^4)$. For a simple calculation we define the function $V$ as\footnote{The equation is invariant under the following transformations: $a\to -a$, $b\to -b$, and $L\to -L$.}
\begin{align}
V&=-2r^3V_\mathrm{eff}
\cr
&=(r^3+a^2r+2Ma^2)\left(\mathcal{E}^2-\frac{b^2}{4M^2}\Delta\right)
-(r-2M)\mathcal{L}^2-4Ma\mathcal{E}\mathcal{L}
-\Delta r\left(1-\frac{b \mathcal{L}}{M}\right),
\end{align}
where $\mathcal{E}=E/m$, $\mathcal{L}=L/m$, and the parameter $b=qBM/m$ introduced in Eq.~\eqref{eq:b} measures influence of the magnetic field on charged particle motion. 

Let us discuss the difference between the ISCO radius for geodesic particle motion and that for charged particle motion. The ISCO radius is determined by solving the equations $V=V'=V''=0$. The first and second derivatives of $V$ with respect to $r$ are given by
\begin{align}
V'=(3r^2+a^2)\mathcal{E}^2-\mathcal{L}^2&-\left(3r^2-4Mr+a^2\right)\left(1-\frac{b\mathcal{L}}{M}\right)
\cr
&-\frac{b^2}{4M^2}\left[5r^4-8Mr^3+6a^2r^2+a^2(a^2-4M^2)\right]
\end{align}
and
\begin{align}
V''=6r\mathcal{E}^2-2(3r-2M)\left(1-\frac{b\mathcal{L}}{M}\right)-\frac{b^2r}{M^2}\left(5r^2-6Mr+3a^2\right),
\end{align}
respectively. We solve $V'=0$ and $V''=0$ for $\mathcal{E}$ and $\mathcal{L}$ as
\begin{equation}
\mathcal{L}=-b\left(r-\frac{a^2}{3r}\right)+\sigma\sqrt{\lambda},
\label{eq:L}
\end{equation}
and
\begin{equation}
\mathcal{E}^2=\eta-\sigma\frac{b}{M}\left(1-\frac{2M}{3r}\right)\sqrt{\lambda},
\label{eq:E}
\end{equation}
respectively, where we have defined $\eta$ and $\lambda$ as
\begin{equation}
\eta=\left(1-\frac{2M}{3r}\right)-\frac{b^2}{6}\left[4-\frac{5r^2}{M^2}
-\frac{a^2}{M^2}\left(3-\frac{2M}{r}+\frac{4M^2}{3r^2}\right)\right]
\end{equation}
and
\begin{align}
\lambda
=&2M\left(r-\frac{a^2}{3r}\right)
\cr
&+\frac{b^2}{4M^2}\left[r^2(5r^2-4Mr+4M^2)+\frac23a^2(5r^2-6Mr+2M^2)
+a^4\left(1+\frac{4M^2}{9r^2}\right)\right],
\label{eq:lambda}
\end{align}
respectively, and $\sigma=\pm1$. Finally, we obtain the equation to determine the radius of the ISCO by substituting Eqs.~\eqref{eq:L} and \eqref{eq:E} into $V=0$, i.e.,
\begin{align}
(r^3+a^2r+2Ma^2)\left(\mathcal{E}^2(r)-\frac{b^2\Delta}{4M^2}\right)-(r-2M)\mathcal{L}^2(r)-4Ma\mathcal{E}(r)\mathcal{L}(r)-\Delta r\left(1-\frac{b\mathcal{L}(r)}{M}\right)=0.
\label{eq:ISCOeq}
\end{align}
We identify the root of Eq.~\eqref{eq:ISCOeq} that is the closest to $r_\mathrm{H}$ as the ISCO radius $r_\mathrm{I}$, where we require $r_\mathrm{I}>r_\mathrm{H}$. Then, from Eqs.~\eqref{eq:L} and \eqref{eq:E}, the energy and the angular momentum of the charged particle at the ISCO are given by $\mathcal{E}_\mathrm{I}=\mathcal{E}(r_\mathrm{I})$ and $\mathcal{L}_\mathrm{I} = \mathcal{L}(r_\mathrm{I})$, respectively. We call the ISCO for $\mathcal{L}_\mathrm{I}>0$ a prograde ISCO and the ISCO for $\mathcal{L}_\mathrm{I}<0$ a retrograde ISCO.

\begin{figure}[t]
\begin{center}
\includegraphics[width=12cm,clip]{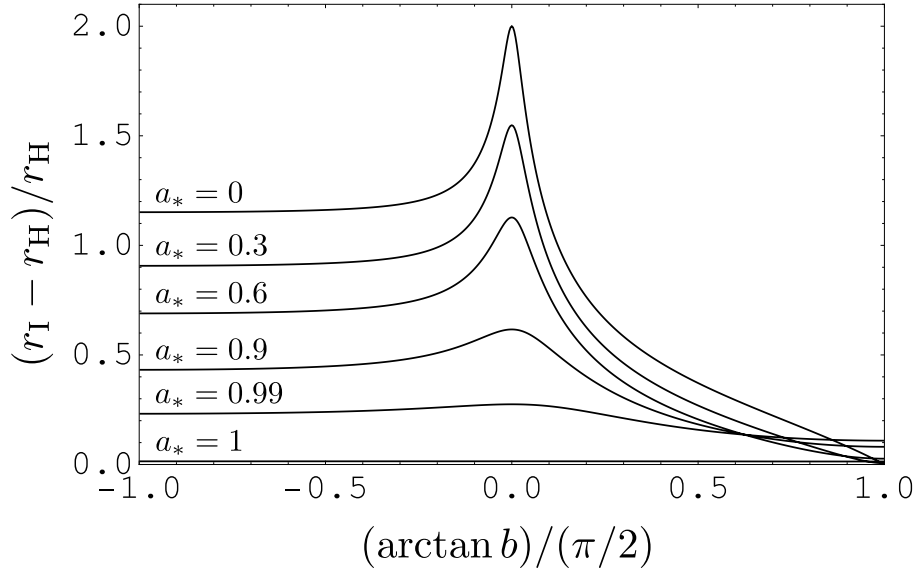}
\end{center}
\caption{The dependence of the ISCO radius $r_\mathrm{I}$ on $b$ for some fixed value of $a_*$.}
\label{fig:ISCO}
\end{figure}

Figure~\ref{fig:ISCO} shows that the dependence of $r_\mathrm{I}$ on $b$ for each fixed value of $a_*$ for the prograde ISCO, where $a_*$ is the dimensionless Kerr parameter defined as $a_*=a/M$. We find that, in the range $0<a_*<1$, $r_\mathrm{I}$ decreases as $|b|$ increases, but approaches a limit value that is larger than $r_\mathrm{H}$ as $|b|\to \infty$. Note, however, that the behavior of $r_\mathrm{I}$ is special in the cases $a_*=0$ and $a_*=1$. In the following subsections, we discuss the dependence of $r_\mathrm{I}$ on $b$ in detail for the case of a nearly maximally rotating magnetized black hole and that of a slowly rotating magnetized black hole. In particular, since the limit value of $r_\mathrm{I}$ as $b\to\infty$ is closer to $r_\mathrm{H}$ than the case of negative $b$, which may cause the fine-tuning of the angular momentum more effectively as mentioned above, we concentrate on the case of nonnegative $b$. In what follows, we assume $a\geq0$ and $b\geq0$.

\subsection{ISCO of a charged particle around a nearly maximally rotating magnetized black hole}
For a nearly maximally rotating ($a_*\simeq1$) black hole with a magnetic field, by solving Eq.~\eqref{eq:ISCOeq} approximately we obtain the radius of the ISCO as
\begin{align}
r_\mathrm{I}/M
=1+2^{2/3}(1-a_*)^{1/3}
+\frac{7+b^2(5-8b^2-6b\sqrt{3+4b^2})}{2^{5/3}(1+b^2)^2}(1-a_*)^{2/3}
+O(1-a_*).
\label{eq:rI(3-A)}
\end{align}
The effect of $b$ appears in the terms of orders higher than $(1-a_*)^{1/3}$. In the maximally rotating black hole case, i.e., $a_*=1$, $r_\mathrm{I}$ coincides with $r_\mathrm{H}$ for any $b$. Then the energy and the angular momentum of the charged particle orbiting the ISCO are given from Eqs.~\eqref{eq:L}--\eqref{eq:lambda} by
\begin{align}
\mathcal{E}_\mathrm{I}
&=\frac{\sqrt{3+4b^2}-b}{3}+\frac{2^{2/3}(\sqrt{3+4b^2}-b)^2}{3\sqrt{3+4b^2}}(1-a_*)^{1/3}
\cr
&\quad+\frac{45-76b^4+4b(3+4b^2)\sqrt{3+4b^2}}{6~2^{2/3}(3+4b^2)^{3/2}}(1-a_*)^{2/3}+O(1-a_*)
\label{eq:E_I(3-A)}
\end{align}
and
\begin{align}
\mathcal{L}_\mathrm{I}/M
&=\frac{2(\sqrt{3+4b^2}-b)}{3}+\frac{2^{5/3}(\sqrt{3+4b^2}-b)^2}{3\sqrt{3+4b^2}}(1-a_*)^{1/3}
\cr
&\quad+\frac{9+2b\left(72b+86b^3-5(3+4b^2)\sqrt{3+4b^2}\right)}{3~2^{2/3}(3+4b^2)^{3/2}}(1-a_*)^{2/3}+O(1-a_*),
\label{eq:L_I(3-A)}
\end{align}
respectively. In the case $b=0$, Eqs.~\eqref{eq:rI(3-A)}--\eqref{eq:L_I(3-A)} reproduce the results of a nearly maximally rotating black hole case, which are seen in Ref.~\cite{Harada:2010yv}.

Let us see the limiting behaviors of Eqs.~\eqref{eq:rI(3-A)}--\eqref{eq:L_I(3-A)}. If we consider the case $b\ll 1$, $r_\mathrm{I}$ behaves as
\begin{align}
r_\mathrm{I}/M=1+2^{2/3}(1-a_*)^{1/3}
+\frac74~2^{1/3}\left(1-\frac97b^2+O(b^3)\right)(1-a_*)^{2/3}+O(1-a_*).
\end{align}
This shows that $r_\mathrm{I}$ becomes smaller by introducing small positive $b$. Then $\mathcal{E}_\mathrm{I}$ and $\mathcal{L}_\mathrm{I}$ become
\begin{align}
\mathcal{E}_\mathrm{I}
&=\frac{1}{\sqrt3}\left(1-\frac{1}{\sqrt3}b+O(b^2)\right)+\frac{2^{2/3}}{\sqrt3}\left(1-\frac{2}{\sqrt3}b+O(b^2)\right)(1-a_*)^{1/3}
\cr&\quad-\frac{5}{4\sqrt3}~2^{1/3}\left(1+\frac{4}{5\sqrt3}b+O(b^2)\right)(1-a_*)^{2/3}+O(1-a_*)
\end{align}
and
\begin{align}
\mathcal{L}_\mathrm{I}/M
&=\frac{2}{\sqrt3}\left(1-\frac{1}{\sqrt3}b+O(b^2)\right)+\frac{4}{2^{1/3}\sqrt3}\left(1-\frac{2}{\sqrt3}b+O(b^2)\right)(1-a_*)^{1/3}
\cr&\quad+\frac{1}{2^{2/3}\sqrt3}\left(1-\frac{10}{\sqrt3}b+O(b^2)\right)(1-a_*)^{2/3}+O(1-a_*),
\end{align}
respectively. Hence, we find that both $\mathcal{E}_\mathrm{I}$ and $\mathcal{L}_\mathrm{I}$ take smaller values by introducing small positive $b$.

We are also interested in the case $b\gg1$. In this case $r_\mathrm{I}$ behaves as
\begin{align}
r_\mathrm{I}/M
=1+2^{2/3}(1-a_*)^{1/3}-5~2^{1/3}\left(1-\frac{81}{40}b^{-2}+O(b^{-4})\right)(1-a_*)^{2/3}+O(1-a_*).
\end{align}
Then $\mathcal{E}_\mathrm{I}$ and $\mathcal{L}_\mathrm{I}$ behave as
\begin{align}
\mathcal{E}_\mathrm{I}
=\frac{1}{3}b\left(1+\frac{1}{2^{1/3}}(1-a_*)^{1/3}+\frac{11}{4~2^{2/3}}(1-a_*)^{2/3}+O(1-a_*)\right)+O(b^{-1})
\end{align}
and
\begin{align}
\mathcal{L}_\mathrm{I}/M=\frac{2}{3}b\left(1+\frac{1}{2^{1/3}}(1-a_*)^{1/3}+\frac{23}{4~2^{2/3}}(1-a_*)^{2/3}+O(1-a_*)\right)+O(b^{-1}),
\end{align}
respectively. Thus, we find that both the energy and the angular momentum of a charged particle at the ISCO become infinite as $b$ goes to infinity.

\subsection{ISCO of a charged particle around a slowly rotating magnetized black hole}
In this subsection, we discuss the dependence of $r_\mathrm{I}$, $\mathcal{E}_\mathrm{I}$, and $\mathcal{L}_\mathrm{I}$ on $a_*$ and $b$ in the case of a slowly rotating magnetized black hole. In Sec.~\ref{sec:3-B-1}, we discuss the case of magnetized Schwarzschild black holes so that we review Frolov's result in the next section. In Secs.~\ref{sec:3-B-1} and \ref{sec:3-B-3}, we discuss the two cases $1\gg b^{-1}\gg a_*$ and $1\gg a_*\gg b^{-1}$ separately because the behavior of $r_\mathrm{I}$, $\mathcal{E}_\mathrm{I}$, and $\mathcal{L}_\mathrm{I}$ is extensively different.

\subsubsection{$a_*=0$ and $b^{-1}\ll 1$}
\label{sec:3-B-1}
Let us consider charged particle motion at the ISCO around a magnetized Schwarzschild black hole. By solving Eq.~\eqref{eq:ISCOeq} with $a=0$ and $b^{-1}\ll 1$, we obtain the asymptotic form of $r_\mathrm{I}$ as
\begin{align}
r_\mathrm{I}/M=2+\frac{2}{\sqrt{3}b}-\frac{8}{9b^2}+O(b^{-3}).
\end{align}
This result coincides with the result given in Ref.~\cite{Frolov:2011ea}. From Eqs.~\eqref{eq:E} and \eqref{eq:L}, $\mathcal{E}_\mathrm{I}$ and $\mathcal{L}_\mathrm{I}$ take the forms
\begin{align}
\mathcal{E}_\mathrm{I}=\frac{2}{3^{3/4}\sqrt{b}}+O(b^{-3/2})
\label{eq:EI(3-B-1)}
\end{align}
and
\begin{align}
\mathcal{L}_\mathrm{I}/M=2b+2\sqrt{3}+O(b^{-1}),
\label{eq:LI(3-B-1)}
\end{align}
respectively. Therefore, the energy of a charged particle at the ISCO approaches zero and the angular momentum takes an arbitrarily high value as $b$ goes to infinity.

\subsubsection{$1\gg b^{-1}\gg a_*$}
\label{sec:3-B-2}
Let us consider charged particle motion at the ISCO around a slowly rotating magnetized black hole in the case $1\gg b^{-1}\gg a_*$. Then the ISCO radius becomes
\begin{align}
r_\mathrm{I}/M=2+\frac{2}{\sqrt{3}b}-\frac{8}{9b^2}+O(b^{-3})
+\left(-\frac{2}{3^{1/4}\sqrt{b}}+O(b^{-3/2})\right)a_*+O(a_*^2).
\label{eq:rI(3-B-2)}
\end{align}
Then $\mathcal{E}_\mathrm{I}$ and $\mathcal{L}_\mathrm{I}$ take the forms
\begin{align}
\mathcal{E}_\mathrm{I}=\frac{2}{3^{3/4}\sqrt{b}}+O(b^{-3/2})+\left(\frac{b}{2}+O(b^0)\right)a_*+O(a_*^2)
\label{eq:EI(3-B-2)}
\end{align}
and
\begin{align}
\mathcal{L}_\mathrm{I}/M=2b+2\sqrt{3}+O(b^{-1})+\left(-2~3^{3/4}\sqrt{b}+O(b^{-1/2})\right)a_*+O(a_*^2),
\label{eq:LI(3-B-2)}
\end{align}
respectively. The leading terms in Eqs.~\eqref{eq:EI(3-B-2)} and \eqref{eq:LI(3-B-2)} reproduce Eqs.~\eqref{eq:EI(3-B-1)} and \eqref{eq:LI(3-B-1)}, respectively. In the next leading terms, we have found the correction of $\mathcal{E}_\mathrm{I}$ and $\mathcal{L}_\mathrm{I}$ by the black hole rotation.

\subsubsection{$1\gg a_*\gg b^{-1}$}
\label{sec:3-B-3}
In the case $1\gg a_*\gg b^{-1}$, $r_\mathrm{I}$ is of the form
\begin{align}
r_\mathrm{I}=2-\frac{3}{8}a_*^2+O(a_*^4)+\left(\frac{a_*^{2/3}}{3^{2/3}}+O(a_*^{8/3})\right)b^{-2/3}+O(b^{-4/3}).
\label{eq:rI(3-B-3)}
\end{align}
Then $\mathcal{E}_\mathrm{I}$ and $\mathcal{L}_\mathrm{I}$ at $r=r_\mathrm{I}$ take the values
\begin{align}
\mathcal{E}_\mathrm{I}=\left(\frac{a_*}{2}-\frac{a_*^3}{32}+O(a_*^5)\right)b
+\left(\frac{a_*^{5/3}}{8~3^{2/3}}+O(a_*^{11/3})\right)b^{1/3}+O(b^{-1/3}),
\label{eq:EI(3-B-3)}
\end{align}
and
\begin{align}
\mathcal{L}_\mathrm{I}/M=\left(2-\frac{3a_*^2}{4}+O(a_*^4)\right)b
+\left(3^{1/3}a_*^{2/3}+O(a_*^{8/3})\right)b^{1/3}+O(b^{-1/3}),
\label{eq:LI(3-B-3)}
\end{align}
respectively. Therefore, unlike in the previous case $1\gg b^{-1}\gg a_*$, both $\mathcal{E}_\mathrm{I}$ and $\mathcal{L}_\mathrm{I}$ go to infinity as $b$ goes to infinity.

In both cases $1\gg b^{-1}\gg a_*$ and $1\gg a_*\gg b^{-1}$, the ISCO radius $r_\mathrm{I}$ given by Eq.~\eqref{eq:rI(3-B-2)} or Eq.~\eqref{eq:rI(3-B-3)} takes a value larger than the horizon radius $r_\mathrm{H}$ since the horizon radius behaves as
\begin{align}
r_\mathrm{H}/M=2-\frac{1}{2}a_*^2+O(a_*^4).
\end{align}
We can say that if the black hole has a small but finite spin the location of the ISCO does not coincide with that of the event horizon even in the limit $b\to \infty$ in contrast to the case of the Schwarzschild black hole.

\section{Evaluation of Center-of-Mass Energy}
\label{sec:4}
In this section we discuss the effect of the magnetic field given in Eqs.~\eqref{eq:uniform magnetic field} on the CM energy of charged particle collisions around a Kerr black hole. Let us obtain the CM energies that are calculated for two different processes in the present section.

The first is the CM energy of charged particle collisions near the horizon; charged particle-1 with $\mathcal{E}_1=\mathcal{E}_\mathrm{I}$ and $\mathcal{L}_1=\mathcal{L}_\mathrm{I}$ plunges from the ISCO to the horizon and collides with generic particle-2. For the magnetic field given in Eqs.~\eqref{eq:uniform magnetic field}, the formula for the CM energy \eqref{eq:Ecm^H} in the near horizon limit is given in the form
\begin{align}
\frac{E_\mathrm{cm}^\mathrm{H}}{\sqrt{m_1m_2}}
=\sqrt{\frac{(m_1+m_2)^2}{m_1m_2}
+\frac{\left[(\mathcal{E}_2\mathcal{L}_1-\mathcal{E}_1\mathcal{L}_2)/(2M)-(b_1\hat{\mathcal{I}}_2-b_2\hat{\mathcal{I}}_1)\right]^2+(\hat{\mathcal{I}}_1-\hat{\mathcal{I}}_2)^2}
{\hat{\mathcal{I}}_1\hat{\mathcal{I}}_2}},
\label{eq:Ecm^H(uniform magnetic field)}
\end{align}
where ${\hat{\mathcal{I}}}_s$ defined in Eq.~\eqref{eq:I_s} takes the form $\hat{\mathcal{I}}_s=\mathcal{E}_s-\Omega_\mathrm{H}\mathcal{L}_s$ because $\Phi_\mathrm{H}$ becomes zero in this case, and $b_s$ denotes the parameter $b$ defined in Eq.~\eqref{eq:b} for particle-$s$.

The second is the case where generic particle-2 collides with charged particle-1 orbiting $r=r_\mathrm{I}$ with $\mathcal{E}_1=\mathcal{E}_\mathrm{I}$ and $\mathcal{L}_1=\mathcal{L}_\mathrm{I}$. From Eq.~\eqref{eq:Ecm^2} evaluated at $r=r_\mathrm{I}$ under the condition $\mathcal{R}_1=0$, the CM energy $E_\mathrm{cm}^\mathrm{I}$ is given by
\begin{align}
\frac{E_\mathrm{cm}^\mathrm{I}}{\sqrt{m_1m_2}}
=&
\Bigg[\frac{m_1^2+m_2^2}{m_1m_2}-\frac{2}{r_\mathrm{I}^2}
\left(\mathcal{L}_\mathrm{I}-a\mathcal{E}_\mathrm{I}-\frac{r_\mathrm{I}^2+a^2}{2M}b_1\right)
\left(\mathcal{L}_2-a\mathcal{E}_2-\frac{r_\mathrm{I}^2+a^2}{2M}b_2\right)
\cr
&+\frac{2}{r_\mathrm{I}^2\Delta_\mathrm{I}}\left((r_\mathrm{I}^2+a^2)\mathcal{E}_\mathrm{I}-a\mathcal{L}_\mathrm{I}+\frac{a\Delta_\mathrm{I}}{2M}b_1\right)
\left((r_\mathrm{I}^2+a^2)\mathcal{E}_2-a\mathcal{L}_2+\frac{a\Delta_\mathrm{I}}{2M}b_2\right)
\Bigg]^{1/2},
\label{eq:Ecm^I(uniform magnetic field)}
\end{align}
where $\Delta_\mathrm{I}=r_\mathrm{I}^2+a^2-2Mr_\mathrm{I}$.

In the following subsections, we evaluate both CM energies, $E_\mathrm{cm}^\mathrm{H}$ and $E_\mathrm{cm}^\mathrm{I}$. We consider the case of a nearly maximally rotating magnetized black hole in Sec.~\ref{sec:4-A} and the case of a slowly rotating magnetized black hole in Sec.~\ref{sec:4-B}.

\subsection{Nearly maximally rotating magnetized black hole}
\label{sec:4-A}
In this subsection, let us evaluate the CM energies $E_\mathrm{cm}^\mathrm{H}$ and $E_\mathrm{cm}^\mathrm{I}$ in a nearly maximally rotating magnetized black hole spacetime. In particular, we focus on the effect of the magnetic field on the BSW process.

Firstly, we estimate $E_\mathrm{cm}^\mathrm{H}$ approximately. Substituting Eqs.~\eqref{eq:E_I(3-A)} and \eqref{eq:L_I(3-A)} into Eq.~\eqref{eq:Ecm^H(uniform magnetic field)}, we obtain
\begin{align}
\frac{E_\mathrm{cm}^\mathrm{H}}{\sqrt{m_1m_2}}
\simeq F(b_1)\frac{2^{1/2}\sqrt{2\mathcal{E}_2-\mathcal{L}_2/M}}{3^{1/4}(1-a_*^2)^{1/4}},
\label{eq:Ecm^H(4-A)}
\end{align}
where the function $F$ is defined by
\begin{align}
F(b)=\frac{\sqrt{-b+\sqrt{3+4b^2}}}{3^{1/4}},
\label{eq:F}
\end{align}
which shows the deviation of $E_\mathrm{cm}^\mathrm{H}$ from the case $b=0$ and behaves as $F(b)\simeq1-b/(2\sqrt{3})$ for $b\ll 1$. Equation \eqref{eq:Ecm^H(4-A)} shows that $E_\mathrm{cm}^\mathrm{H}$ becomes arbitrarily large as $a_*$ goes to unity. This is caused by the fine-tuning of the angular momentum, $\mathcal{I}_1=0$, for the charged particle collision, which is given by Eqs.~\eqref{eq:I_s}, \eqref{eq:E_I(3-A)}, and \eqref{eq:L_I(3-A)}. The result means that the BSW process can occur in the case of the charged particle collisions in the nearly maximally rotating magnetized black hole. Therefore, we conclude that the BSW process is stable against the effect of the magnetic field.

In the case $b\gg 1$, which is relevant to astrophysics, $F(b)$ behaves as $F(b) \simeq 3^{-1/4}\sqrt{b}$. Therefore, the CM energy can be arbitrarily high as the magnetic field is arbitrarily strong, even though the black hole spin is not nearly maximal. Note, however, that we should distinguish this effect from the BSW process because in this case either or both of energies of the charged particles plunging from the ISCO must be arbitrarily large before the collision in the limit $b\to \infty$.

Secondly, we estimate $E_\mathrm{cm}^\mathrm{I}$ approximately. Substituting Eqs.~\eqref{eq:E_I(3-A)} and \eqref{eq:L_I(3-A)} into Eq.~\eqref{eq:Ecm^I(uniform magnetic field)}, we have
\begin{align}
\frac{E_\mathrm{cm}^\mathrm{I}}{\sqrt{m_1m_2}}
\simeq F(b_1)~\frac{2^{5/6}\sqrt{2\mathcal{E}_2-\mathcal{L}_2/M}}{3^{1/4}(1-a_*^2)^{1/6}},
\label{eq:Ecm^I(4-A)}
\end{align} 
where this estimate also includes the same function $F(b_1)$ defined in Eq.~\eqref{eq:F}. For $b_1=0$, Eq.~\eqref{eq:Ecm^I(4-A)} reproduces Eq.~(5.1) in Ref.~\cite{Harada:2010yv}. We find that $E_\mathrm{cm}^\mathrm{I}$ also becomes arbitrarily large as $a_*$ goes to unity. The result also means that the BSW process can occur in the case of the charged particle collisions in the nearly maximally rotating magnetized black hole.

Let us evaluate typical values of Eqs.~\eqref{eq:Ecm^H(4-A)} and \eqref{eq:Ecm^I(4-A)}. For a collision of an electron as particle-$1$ and a hydrogen atom
 as particle-$2$ around a stellar mass black hole with $M=10M_{\odot}$ and $B=10^8\,\mathrm{Gauss}$, $F(b_1)$ is estimated to be $7.0\times10^5$, where $b_1\simeq8.6\times 10^{11}$. If we take Thorne's limit, $a_*=0.998$, the ratio of $E_\mathrm{cm}^\mathrm{H}$ to the rest electron mass $m_\mathrm{e}$ becomes $E_\mathrm{cm}^\mathrm{H}/m_\mathrm{e} \simeq 1.3\times10^8$, where we have assumed $\sqrt{2\mathcal{E}_2-\mathcal{L}_2/M}=1$. Namely, $E_\mathrm{cm}^\mathrm{H}$ is much greater than $m_\mathrm{e}$. The ratio $E_\mathrm{cm}^\mathrm{I}/m_\mathrm{e}$ also takes a large value $E_\mathrm{cm}^\mathrm{I}/m_\mathrm{e}\simeq1.0 \times 10^8$ under the same choice of the parameters. Namely, $E_\mathrm{cm}^\mathrm{H}$ and $E_\mathrm{cm}^\mathrm{I}$ are estimated to be $\sim66\,\mathrm{TeV}$ and $\sim 52\,\mathrm{TeV}$, respectively. Furthermore, Table~\ref{table:1} shows that the CM energies of various colliding particle pairs in the same situation. Therefore, we find that highly relativistic collisions occur near the horizon of a nearly maximally rotating magnetized black hole. Note that the high collision energy here is caused by the requirement of the high initial energy, $\mathcal{E}_1\propto b_1$, for the charged particles to orbit the ISCO.
\begin{center}
\begin{table}
\begin{tabular}{ccc}
\hline\hline
\hspace{5mm} colliding particles \hspace{5mm}&
\hspace{5mm} $E_\mathrm{cm}^\mathrm{H}$ (TeV)\hspace{5mm}&
\hspace{5mm} $E_\mathrm{cm}^\mathrm{I}$ (TeV) \hspace{5mm}
\\ \hline
$\mathrm{e}$-$\mathrm{H}$&$6.6\times10^1$&$5.2\times10^1$\\
$\mathrm{p}$-$\mathrm{H}$&$6.6\times10^1$&$5.2\times10^1$\\
$\mathrm{Fe}^{26+}$-$\mathrm{H}$&$3.4\times10^2$&$2.7\times10^2$\\
$\mathrm{Fe}^{26+}$-$\mathrm{Fe}$&$2.5\times10^3$&$2.0\times10^3$\\
\hline\hline
\end{tabular}
\caption{The CM energies for collisions of an electron and a hydrogen atom ($\mathrm{e}$-$\mathrm{H}$), a proton and a hydrogen atom ($\mathrm{p}$-$\mathrm{H}$), an iron nucleus and a hydrogen atom ($\mathrm{Fe}^{26+}$-$\mathrm{H}$), and an iron nucleus and an iron atom ($\mathrm{Fe}^{26+}$-$\mathrm{Fe}$), where we have used $a_*=0.998$, $M=10M_{\odot}$, $B=10^8\,\mathrm{Gauss}$, $\sqrt{2\mathcal{E}_2-\mathcal{L}_2/M}=1$, $m_\mathrm{e}=0.511\,\mathrm{MeV}$, $m_\mathrm{p}=m_\mathrm{H}=0.938\,\mathrm{GeV}$, and $m_\mathrm{Fe}=m_{\mathrm{Fe}^{26+}}=55.9m_\mathrm{p}$.}
\label{table:1}
\end{table}
\end{center}

\subsection{Slowly rotating magnetized black hole}
\label{sec:4-B}
In this subsection, we focus on the case of a slowly rotating magnetized black hole. Let us evaluate $E_\mathrm{cm}^\mathrm{H}$ and $E_\mathrm{cm}^\mathrm{I}$ in three cases: $a_*=0$ and $b^{-1}\ll 1$ in Sec.~\ref{sec:4-B-1}; $1\gg b_s^{-1} \gg a_*$ in Sec.~\ref{sec:4-B-2}; $1\gg a_* \gg b_s^{-1}$ in Sec.~\ref{sec:4-B-3}.

\subsubsection{$a_*=0$ and $b_1^{-1}\ll 1$}
\label{sec:4-B-1}
In the case $a_*=0$ and $b_1^{-1}\ll 1$, which is the case of a magnetized Schwarzschild black hole, we estimate Eq.~\eqref{eq:Ecm^H(uniform magnetic field)} by using \eqref{eq:EI(3-B-1)} and \eqref{eq:LI(3-B-1)}. Then we have
\begin{align}
\frac{E_\mathrm{cm}^\mathrm{H}}{\sqrt{m_1m_2}}
=3^{3/8}\sqrt{2\mathcal{E}_2}~b_1^{1/4}+O(b_1^{-1/4}).
\label{eq:Ecm^H(4-B-3)}
\end{align}
This result shows that $E_\mathrm{cm}^\mathrm{H}$ becomes arbitrarily large as $b_1$ goes to infinity. Note, however, that the colliding particles have finite energies before the collision. We call the collision with an arbitrarily high CM energy of charged particles as $b_1\to \infty$ with initial finite energies Frolov's process. In this case, $\hat{\mathcal{I}}_1$ approaches zero, which is similar to that in the BSW process in a maximally rotating black hole with no magnetic field. Consequently, we find that the arbitrarily high energy collisions of particles with finite energies are realized in the magnetized Schwarzschild black hole. 

On the other hand, $E_\mathrm{cm}^\mathrm{I}$ behaves as
\begin{align}
\frac{E_\mathrm{cm}^\mathrm{I}}{\sqrt{m_1m_2}}
=\frac{2\sqrt{\mathcal{E}_2}}{3^{1/8}}~b_1^{1/4}+O(b_1^{-1/4}).
\label{eq:Ecm^I(4-B-1)}
\end{align}
This result also shows that $E_\mathrm{cm}^\mathrm{I}$ becomes arbitrarily large as $b_1$ goes to infinity, where the initial energies of colliding particles are finite. Since Eq.~\eqref{eq:Ecm^I(4-B-1)} reproduces Frolov's result Eq.~(42) in Ref.~\cite{Frolov:2011ea}, we find that Frolov's process here is the original one. 

Note that in both cases the angular momentum of the charged particle orbiting the ISCO becomes infinite as $b$ goes to infinity. However, this does not immediately mean that the angular momentum measured by a local observer, $\pi_\phi$, becomes infinite. In fact, $\pi_\phi$ approaches zero as $b$ goes to infinity in these cases.

\subsubsection{$1\gg b_1^{-1} \gg a_*$}
\label{sec:4-B-2}
In the case $1\gg b_1^{-1} \gg a_*$, which includes the Schwarzschild limit, substituting Eqs.~\eqref{eq:EI(3-B-2)} and \eqref{eq:LI(3-B-2)} into Eq.~\eqref{eq:Ecm^H(uniform magnetic field)}, we have
\begin{align}
\frac{E_\mathrm{cm}^\mathrm{H}}{\sqrt{m_1m_2}}
=3^{3/8}\sqrt{2\mathcal{E}_2}~b_1^{1/4}+O(b_1^{-1/4})
+\left(\frac{3^{3/8}(\mathcal{L}_2/M-4b_2)}{4\sqrt{2\mathcal{E}_2}}b_1^{1/4}+O(b_1^{-1/4})\right)a_*+O(a_*^2).
\label{eq:Ecm^H(4-B-2)}
\end{align}
The leading term of Eq.~\eqref{eq:Ecm^H(4-B-2)} reproduces Eq.~\eqref{eq:Ecm^H(4-B-3)}. As seen from the next leading term, the rotation of the black hole provides a small correction to Eq.~\eqref{eq:Ecm^H(4-B-3)}. Note that $E_\mathrm{cm}^\mathrm{H}$ becomes arbitrarily large as $b_1$ goes to infinity only in the case of the magnetized Schwarzschild black hole because we cannot discuss $b_1 \to \infty$ for a nonzero $a_*$ in the present parameter regime. Frolov's process in the case of a slowly rotating magnetized black hole will be discussed in the next section.

On the other hand, $E_\mathrm{cm}^\mathrm{I}$ behaves as
\begin{align}
\frac{E_\mathrm{cm}^\mathrm{I}}{\sqrt{m_1m_2}}
=\frac{2\sqrt{\mathcal{E}_2}}{3^{1/8}}~b_1^{1/4}+O(b_1^{-1/4})
+\left(\frac{3^{1/8}\sqrt{\mathcal{E}_2}}{4}~b_1^{3/4}+O(b_1^{1/4})\right)a_*+O(a_*^2).
\label{eq:Ecm^I(4-B-2)}
\end{align}
The leading term of Eq.~\eqref{eq:Ecm^I(4-B-2)} also reproduces Eq.~\eqref{eq:Ecm^I(4-B-1)}. As seen the correction in the next leading term, we find that small rotation of the black hole enhances the value of $E_\mathrm{cm}^\mathrm{I}$. As in the case of Eq.~\eqref{eq:Ecm^H(4-B-2)}, Eq.~\eqref{eq:Ecm^I(4-B-2)} becomes arbitrarily large as $b_1$ goes to infinity only in the limit of the magnetized Schwarzschild black hole.

\subsubsection{$1\gg a_* \gg b_1^{-1}$}
\label{sec:4-B-3}
We turn our attention to the case $1\gg a_* \gg b_1^{-1}$, which corresponds to typical astrophysical situations. Substitution of Eqs.~\eqref{eq:EI(3-B-3)} and \eqref{eq:LI(3-B-3)} into Eq.~\eqref{eq:Ecm^H(uniform magnetic field)} yields
\begin{align}
\frac{E_\mathrm{cm}^\mathrm{H}}{\sqrt{m_1m_2}}
=\left(\frac{3\sqrt{\mathcal{E}_2}}{\sqrt{2}}a_*^{1/2}+O(a_*^{3/2})\right)b_1^{1/2}
+\left(3^{4/3}\sqrt{2\mathcal{E}_2}~a_*^{-5/6}+O(a_*^{1/6})\right)b_1^{-1/6}
+O(b_1^{-5/6}).
\label{eq:Ecm^H(3-B-3)}
\end{align}
On the other hand, from Eq.~\eqref{eq:Ecm^I(uniform magnetic field)} with Eqs.~\eqref{eq:EI(3-B-3)} and \eqref{eq:LI(3-B-3)}, $E_\mathrm{cm}^\mathrm{I}$ becomes
\begin{align}
\frac{E_\mathrm{cm}^\mathrm{I}}{\sqrt{m_1m_2}}
=\left(2\sqrt{\mathcal{E}_2}~a_*^{1/2}+O(a_*^{3/2})\right)b_1^{1/2}
+\left(-\frac{7\sqrt{\mathcal{E}_2}}{3^{2/3}}a_*^{-5/6}+O(a_*^{1/6})\right)b_1^{-1/6}
+O(b_1^{-1/2}).
\label{eq:Ecm^I(3-B-3)}
\end{align}
The both results show that $E_\mathrm{cm}^\mathrm{H}$ and $E_\mathrm{cm}^\mathrm{I}$ become arbitrarily large as $b_1$ goes to infinity. The power of $b_1$ in Eqs.~\eqref{eq:Ecm^H(3-B-3)} and \eqref{eq:Ecm^I(3-B-3)} directly reflects that in Eq.~\eqref{eq:EI(3-B-3)}. Note, however, that arbitrarily high initial energy of the colliding charged particle is required to orbit the ISCO in order that we obtain arbitrarily large $E_\mathrm{cm}^\mathrm{H}$ or $E_\mathrm{cm}^\mathrm{I}$. Therefore, we conclude that Frolov's process does not occur in this case.

As discussed in the previous subsection, we consider the collision of an electron as particle-$1$ and a hydrogen atom as particle-$2$ around a stellar mass black hole with $M=10M_{\odot}$, $a_*=0.01$, and $B=10^8\,\mathrm{Gauss}$, where $b_1\simeq8.6\times10^{11}$. We should note that the estimate of the CM energy by Frolov for the Schwarzschild black hole is not applicable for realistic astrophysical situations because the condition $b^{-1}\gg a_{*}$ would not be satisfied there. Since the parameters satisfy $1\gg a_* \gg b_1^{-1}$, we should adopt Eqs.~\eqref{eq:Ecm^H(3-B-3)} and \eqref{eq:Ecm^I(3-B-3)}. Assuming $\sqrt{\mathcal{E}_2}=1$, we obtain $E_\mathrm{cm}^\mathrm{H}/m_\mathrm{e}\simeq8.4\times10^6$ and $E_\mathrm{cm}^\mathrm{I}/m_\mathrm{e} \simeq7.9\times10^{6}$. Namely, both of the CM energies are $\sim 4\,\mathrm{TeV}$ in order of magnitude. Furthermore, Table~\ref{table:2} shows that the CM energies of various colliding particle pairs in the similar situation. This means that highly relativistic collision can occur near the horizon of a slowly rotating black hole in an astrophysical context. However, the high collision energy is caused by the high initial energy of the charged particle, $\mathcal{E}_1\propto b_1$, to orbit the ISCO. Finally, we conclude that Frolov's process does not occur in the slowly rotating magnetized black hole in the present choice of field configuration.
\begin{center}
\begin{table}
\begin{tabular}{ccc}
\hline\hline
\hspace{5mm}colliding particles\hspace{5mm}&
\hspace{5mm}$E_\mathrm{cm}^\mathrm{H}$ (TeV)\hspace{5mm}&
\hspace{5mm}$E_\mathrm{cm}^\mathrm{I}$ (TeV)\hspace{5mm}
\\ \hline
$\mathrm{e}$-$\mathrm{H}$&$4.3\times10^{1-n}$&$4.1\times10^{1-n}$\\
$\mathrm{p}$-$\mathrm{H}$&$4.3\times10^{1-n}$&$4.1\times10^{1-n}$\\
$\mathrm{Fe}^{26+}$-$\mathrm{H}$&$2.2\times10^{2-n}$&$2.1\times10^{2-n}$\\
$\mathrm{Fe}^{26+}$-$\mathrm{Fe}$&$1.6\times10^{3-n}$&$1.5\times10^{3-n}$\\
\hline \hline
\end{tabular}
\caption{The CM energies for collisions of an electron and a hydrogen atom ($\mathrm{e}$-$\mathrm{H}$), a proton and a hydrogen atom ($\mathrm{p}$-$\mathrm{H}$), an iron nucleus and a hydrogen atom ($\mathrm{Fe}^{26+}$-$\mathrm{H}$), and an iron nucleus and an iron atom ($\mathrm{Fe}^{26+}$-$\mathrm{Fe}$). We have defined $a_*=10^{-2n}$, where $n$ is determined in the range $1\gg a_*\gg b_1^{-1}$ and have used that $M=10M_{\odot}$, $B=10^8\,\mathrm{Gauss}$, $\sqrt{\mathcal{E}_2}=1$, $m_\mathrm{e}=0.511\,\mathrm{MeV}$, $m_\mathrm{p}=m_\mathrm{H}=0.938\,\mathrm{GeV}$, and $m_\mathrm{Fe}=m_{\mathrm{Fe}^{26+}}=55.9 m_\mathrm{p}$.}
\label{table:2}
\end{table}
\end{center}

\section{Conclusion}
In this paper, we have discussed the effect of weak electromagnetic fields on charged particle acceleration by a Kerr black hole. We have obtained the general formula for the CM energy of non-geodesic particle collisions, in particular, charged particles on the equatorial plane in test electromagnetic fields around a Kerr black hole. The CM energy evaluated near the horizon becomes arbitrarily high if either particle is near critical one.

The orbits of charged particles in a magnetized Kerr black hole dramatically deviate from geodesics even for a modest electromagnetic field. In this paper, we have discussed charged particles at the ISCO in the uniform magnetic field around a Kerr black hole. As a result, we have found that the ISCO shifts inward because of the effect of the magnetic field. To see the effect of the magnetic field on charged particle collisions near the horizon, we have derived the modification to the CM energy in two cases: a charged particle plunges from the ISCO to the horizon and collides with another particle; a charged particle orbiting the ISCO collides with another particle at the ISCO.

We have embedded the BSW process and Frolov's process into the particle acceleration by a weakly magnetized Kerr black hole. Firstly, for a nearly maximally rotating magnetized black hole, we have obtained the correction of the CM energy for the BSW effect by the magnetic field. The results show that the BSW process occurs even for the charged particle collisions in the magnetic field around a rotating black hole, which is caused by the fine-tuning of the angular momentum of a charged particle at the ISCO. On the other hand, the CM energy can be arbitrarily high as the magnetic field is arbitrarily strong, even though the black hole spin is not nearly maximal. However, we should distinguish this effect from the BSW process because in the present case the energy of charged particles plunging from the ISCO must be arbitrarily large before the collision.

Secondly, we have discussed charged particle collisions around a slowly rotating magnetized black hole. We have found that high energy charged particle collisions occur even in the case of a non-extremal black hole by the effect of the magnetic field, while the arbitrarily high energy collisions of particles with finite energies are realized only in the magnetized Schwarzschild black hole. Furthermore, in a typical situation of astrophysics, Frolov's process does not occur at least in the present choice of the field configuration because large energy is required to put a charged particle at the ISCO. It is not clear whether such extremely high energy particles are realistic around astrophysical compact objects.

We have estimated the typical values for the CM energy for the collision of a charged particle orbiting around a magnetized Kerr black hole with the parameter values which are realistic in astrophysical stellar mass black holes both for the rapid rotation case and for the slow rotation case. The result is summarized in Tables~\ref{table:1} and \ref{table:2}. We can see that the typical value ($\sim 10\,\mathrm{TeV}$ per nucleon) for the former case $a_{*}= 0.998$ is much greater than that ($\sim 1\,\mathrm{TeV}$ per nucleon) for the latter case $a_{*}=0.01$. From this result, we can conclude that the combination of the large spin and the strong magnetic field of the black hole efficiently accelerates a charged particle orbiting around the black hole. The typical values for the CM energy in the case of a supermassive black hole are almost the same as or slightly larger than those in the case of a stellar mass black hole.

In the present choice of the magnetic field configuration, the electric charge of the spacetime is given by $Q=2MaB$~\cite{Wald:1974np}, which is nonzero for a rotating black hole but typically gives a very small value $\sim 10^{-10}$ for $Q/M$. Note, however, that this field configuration is not unique. For example, to see the pure effect of the magnetic field, we should choose the electromagnetic field that has no electric charge, because the behavior of both the BSW process and Frolov's process may depend on the choice of the field configuration.

\acknowledgments
The authors thank N.~Shibazaki, H.~Ishihara, Y.~Takamori, and the members of the YITP cosmology group for very helpful comments and suggestions. T.I. was supported by Grant-in-Aid for JSPS Fellows No.~11J08747. T.H. was supported by Grant-in-Aid for Scientific Research from the Ministry of Education, Culture, Sports, Science and Technology of Japan [Young Scientists (B) No. 21740190]. M.K. was supported by Grant-in-Aid for JSPS Fellows No.~11J02182.

\appendix
\section{Regularity of $\xi^{a}$ and $\psi^{a}$ at the horizon}
\label{app:A}
Let us consider a coordinate transformation defined by
\begin{align}
dt&=dT-\frac{2Mr}{\Delta}dr,
\label{eq:dt}
\\
d\phi&=d\Phi-\frac{a}{\Delta}dr,
\label{eq:dphi}
\end{align}
from the the Boyer-Lindquist coordinates $(t,r,\theta, \phi)$ to the Kerr-Schild coordinates $(T,r,\theta, \Phi)$. In these coordinates, the metric of the Kerr spacetime \eqref{eq:metric} can be rewritten as
\begin{align}
ds^2&=-\left(1-\frac{2Mr}{\Sigma}\right)dT^2+\frac{4Mr}{\Sigma}dTdr-\frac{4Mr}{\Sigma}a\sin^2\theta dTd\Phi+\left(1+\frac{2Mr}{\Sigma}\right)dr^2+\Sigma d\theta^2
\cr
&\quad+\frac{(r^2+a^2)^2-\Delta a^2\sin^2\theta}{\Sigma}\sin^2\theta d\Phi^2
-2a\sin^2\theta \left(1+\frac{2Mr}{\Sigma}\right)drd\Phi.
\label{Kerr-Schild metric}
\end{align}
We can easily check that the metric \eqref{Kerr-Schild metric} is regular at the black hole horizon $r = r_\mathrm{H}$, i.e., the metric components do not diverge at the horizon and the determinant of the metric components takes a finite negative value at the horizon. Thus, the Kerr-Schild coordinates cover the black hole horizon $r = r_\mathrm{H}$ in the Kerr spacetime, while the Boyer-Lindquist coordinates do not.

From the relation between the Boyer-Lindquist coordinates and the Kerr-Schild coordinates \eqref{eq:dt} and \eqref{eq:dphi}, we can show
\begin{align}
\left(\partial/\partial t\right)^a
&=\left(\partial/\partial T\right)^a,
\\
\left(\partial/\partial \phi\right)^a
&=\left(\partial/\partial \Phi\right)^a.
\end{align}
Since the coordinate bases in the Kerr-Schild coordinates are regular at the horizon $r = r_\mathrm{H}$, we can say that the two vector fields $\xi^a=(\partial/\partial t)^a$ and $\psi^a=(\partial/\partial \phi)^a$ in the Boyer-Lindquist coordinates are also regular at the horizon $r = r_\mathrm{H}$.

\section{Circular orbits near the ISCO}
\label{app:B}
For a particle in the circular orbit to reach the ISCO by emitting its energy and angular momentum, it is necessary that the energy of the circular orbits near the ISCO monotonically decreases as the radius of the circular orbit decreases.
By solving $V=0$ and $V'=0$, we can find energy $E$ and angular momentum $L$ as the function of the radius of the circular orbit $r$. We should check $dE/dr>0$ near the ISCO. In this section, we numerically show that this condition is satisfied for typical examples.

\begin{figure}[!h]
\begin{center}
\begin{picture}(0,0)(0,0)
{\small
\put(55,124){(a)}
\put(246,125){(b)}
\put(142,-21){(c)}
}
\end{picture}
\includegraphics[width=0.4\linewidth,clip]{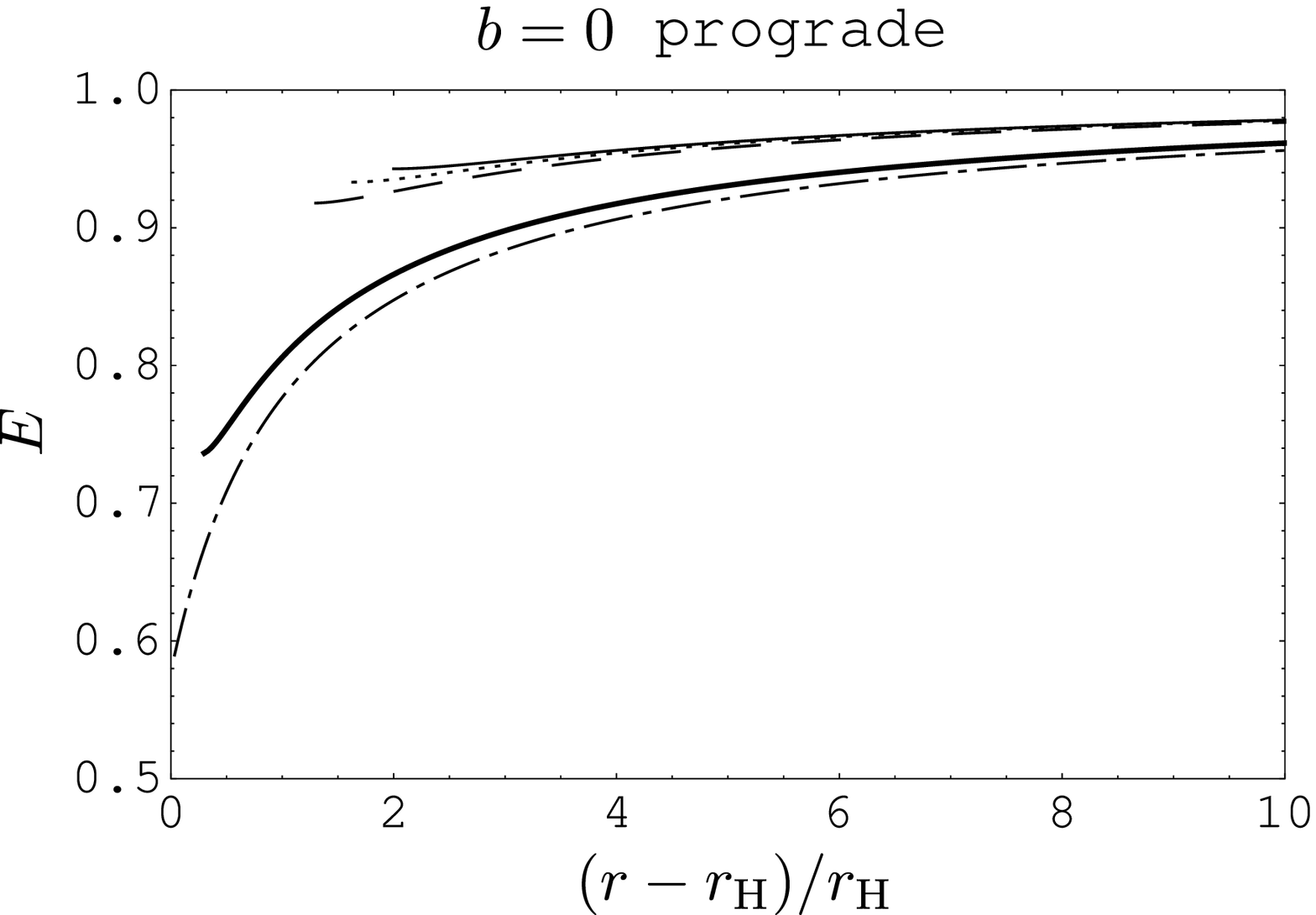}
\includegraphics[width=0.4\linewidth,clip]{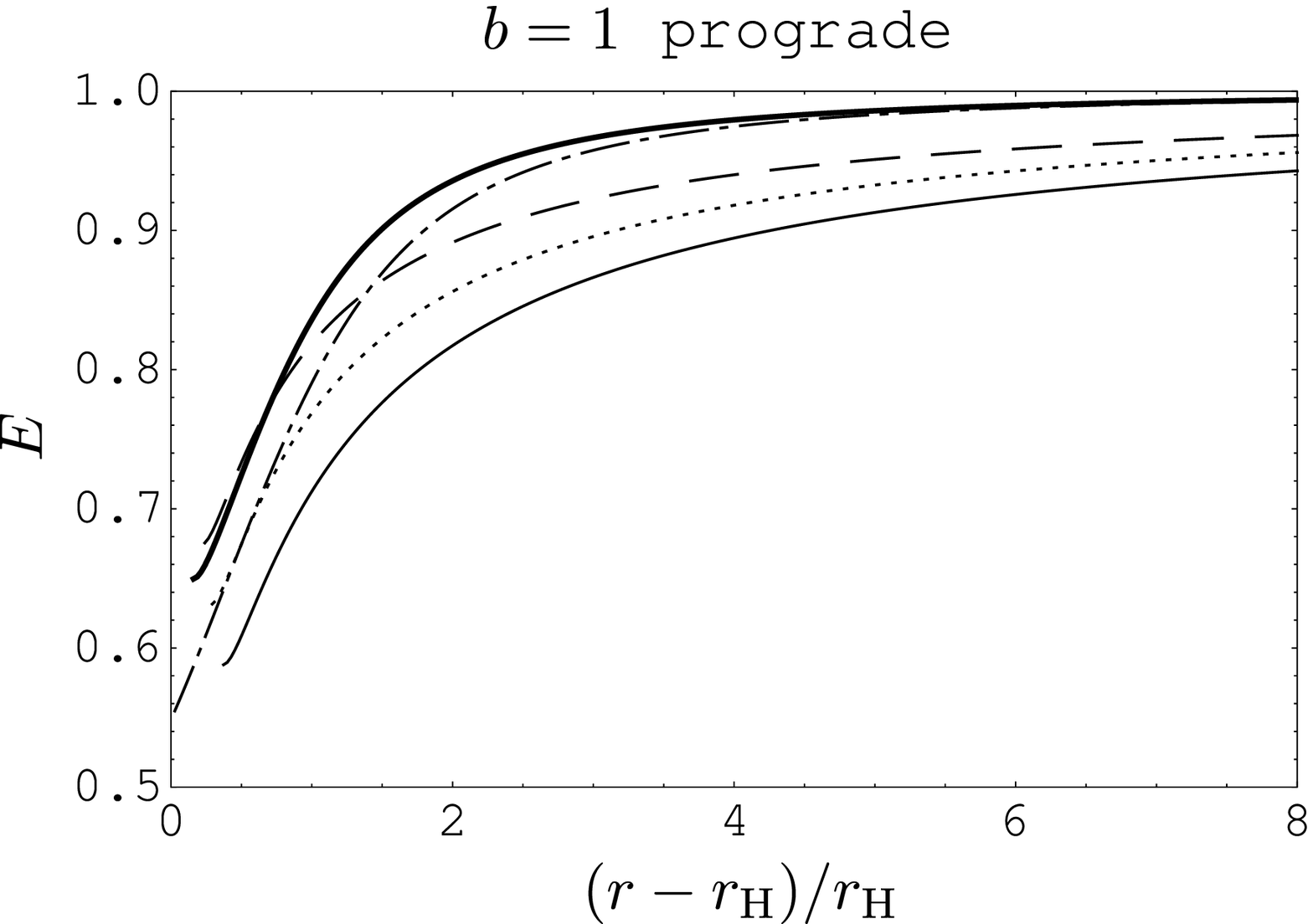}\\
\vspace{5mm}
\includegraphics[width=0.4\linewidth,clip]{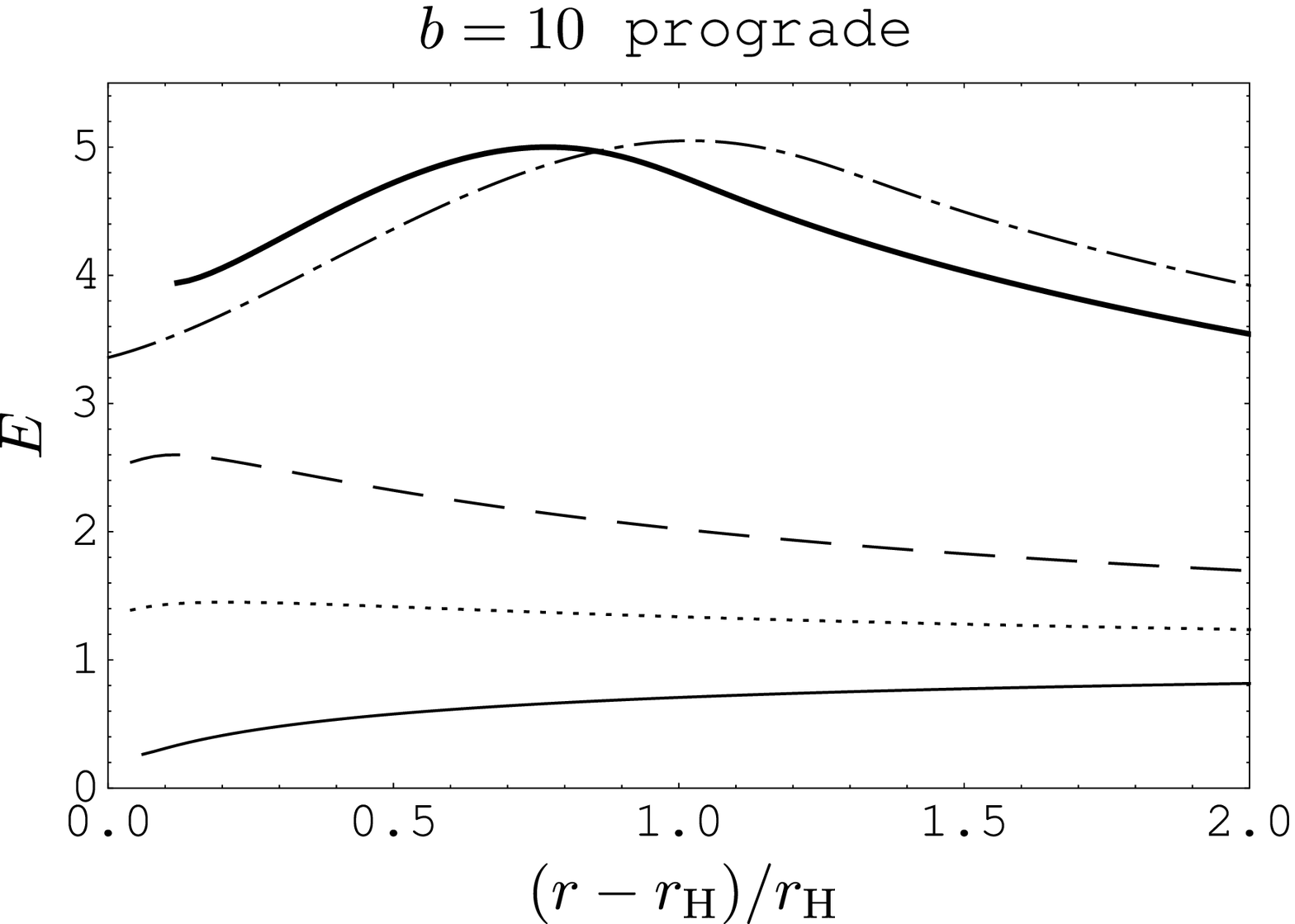}
\end{center}
\caption{
Panels (a), (b), and (c) show the energy of a charged particle as the function of the radius of a circular orbit for $b=0$, $1$, and $10$, respectively. The solid, dotted, dashed, thick, and dot-dashed curves in each panel denote the cases $a=0,~0.25,~0.5,~0.99$, and $1$, respectively.}
\label{fig:circular orbits}
\end{figure}

We plot the energy and the angular momentum for the typical sequences of circular orbits which contain the ISCO in Fig.~\ref{fig:circular orbits}. In this figure, the leftmost point corresponds to the ISCO for each sequence. From this figure, we can see that the condition $dE/dr>0$ is satisfied for the circular orbits near the ISCO, although this is not always satisfied for circular orbits distant from the ISCO. We also note that the function $dE/dr$ takes zero at the ISCO radius.

\end{document}